\documentclass[aps,prb,twocolumn,groupedaddress]{revtex4-1}
\usepackage{graphicx}
\usepackage{bm}
\usepackage{url}
\pdfoutput=1

\newcommand{\Kvec}{\mathbf{K}}
\newcommand{\Gvec}{\mathbf{G}}
\newcommand{\kvec}{\mathbf{k}}
\newcommand{\Rvec}{\mathbf{R}}
\newcommand{\rvec}{\mathbf{r}}
\renewcommand{\Re}{\mathrm{Re}}
\begin{document}

\title{Visualizing influence of point defects on electronic band structure of graphene}

\author{M. Farjam}
\affiliation{School of Nano-Science, Institute for Research in Fundamental
Sciences (IPM), P.O.~Box 19395-5531, Tehran, Iran}

\begin{abstract}
The supercell approach enables us to treat the electronic structure of
defective crystals,
but the calculated energy bands are too complicated
to understand or to compare with angle-resolved photoemission spectra
because of inevitable zone folding.
We discuss how to visualize supercell
band structures more effectively by incorporating in them
unfolded spectral weights and orbital decompositions.
We then apply these ideas to gain a better understanding of the band structure
of graphene containing various types of points defects,
including nitrogen impurity, hydrogen adsorbate, and vacancy defect,
and also the Stone-Wales defect.
\end{abstract}


\maketitle

\section{Introduction}
The electronic band structure of a crystal is an important tool
to understand its electronic, optical, transport and magnetic properties.
Applied to graphene,
the band structure already
shows us many interesting facts about its electronic properties.
\cite{castro-neto2009, wallace1947}
We can see that graphene is a gapless semiconductor,
and its low-energy excitations have linear dispersion,
which implies that they have the properties of massless Dirac fermions.
\citep{novoselov2004, *novoselov2005, *zhang2005}

Defects and impurities have a significant influence on the
electronic properties of semiconductors,
and can be introduced deliberately to tailor their electronic structure.
In graphene, atomic impurities, point and structural defects, edges
and substrates can all modify the electronic structure in important ways.
Two simple impurities are substitutional dopants formed by carbon neighbors
in the periodic table, boron and nitrogen,
which can turn graphene into a p- or n-type semiconductor, respectively.
\cite{usachov2011}
Two other simple defects of a different type are vacancies and hydrogen adsorbates,
which induce midgap states and magnetic moments in graphene.
\cite{yazyev2007}
A small structural defect is the 57-57 defect
known as Stone-Wales (SW) defect.
\cite{stone1986, banhart2011}

A convenient way of treating
point defects is via the supercell approach
within density-functional theory (DFT) and tight-binding method calculations.
The supercell device allows us to use methods
designed for periodic structures
to treat relatively isolated defects.
The work-around has a price: the Brillouin zone (BZ) becomes smaller as
the supercell gets larger,
which results in the folding of the bands.
The complicated bands are difficult to comprehend
or to compare with angle-resolved photoemission spectroscopy (ARPES) experiments.
This often leaves the density of states (DOS)
as the sole option for spectral analysis.

However, there are ways to improve visualization of band structures.
A simplification is made possible by the unfolding method,
which allows plotting effective bands in the larger BZ of the normal system,
i.e., one described in terms of the primitive unit cell.
\citep{allen2013, boykin2005, *boykin2007, ku2010, lee2013, popescu2010, *popescu2012}
Furthermore, an enhancement can be made by incorporating
orbital contributions in both supercell and normal cell band structures.
Orbital decomposition is more commonly used in obtaining
partial density of states (PDOS),
but it is sometimes seen in band structure plots as well.
\cite{rudenko2013}
In this paper, we develop ways of implementing the above ideas
in the visualization of electronic structure data.
We then apply them to the case of graphene containing point defects.
Our results are based on density-functional based tight-binding (DFTB) calculations
for $5\times5$ supercells of graphene containing a nitrogen impurity,
a hydrogen adsorbate, a vacancy, and a SW defect, respectively.

In Sec.~II we describe the computational details,
including DFTB calculations and unfolding procedure.
In Sec.~III we present our results and discussion for perfect graphene
and each type of defect.
In Sec.~IV we summarize our conclusions.

\section{Computational details}
\subsection{DFTB calculations}
We calculate the prerequisite electronic structure data
for the unfolding method by the DFTB+ code,
\cite{[][{. We used the mio-0-1 set of Slater-Koster parameters,
obtained from \url{http://www.dftb.org.}}]aradi2007}
which is an efficient and accurate implementation of the DFTB method.
\cite{elstner1998}
Since the method is based on a small set of non-orthogonal atomic orbitals,
$2s$, $2p_x$, $2p_y$, and $2p_z$,
for carbon and nitrogen, and $2s$ for hydrogen,
it is also quite suitable for the unfolding formula to be described below.

We use $n\times n$ supercells,
which encompass $N_u=n^2$ normal unit cells,
or $2n^2$ honeycomb lattice sites.
This implies for $n=5$ a $2\%$ concentration
of periodically arranged point defects,
which is sufficiently small to consider them as relatively isolated,
but large enough to produce visible effects.
We need to run the code several times.
Having defined a supercell with our choice of defect placed in a tentative position,
in a first calculation we allow the atomic positions to relax.
Then using the optimized coordinates,
we calculate the DOS and PDOS
using a finer $k$-mesh for the desired accuracy.
Finally, we use the well-converged charges
saved from the previous self-consistent calculation
to obtain the data for the band structure along the
given paths in supercell and normal cell Brillouin zones, respectively.

The basic ingredients needed for the unfolding method consist of
eigenvector coefficients and overlap integrals.
Thus concomittant with the band structure calculations,
we instruct the code to save these data in designated files.
To estimate the size of the data,
which may grow quite large,
let $N_b$ be the total number of orbitals in a supercell,
which is also the total number of bands,
and let $N_k$ be the number of k
points used in the calculation of the band structure.
The overlap integrals are represented as a sparse matrix,
\cite{aradi2007}
so their size can only be estimated to be less than
$N_R\times N_b\times N_b\times \mathrm{(size\ of\ a\ real\ number)}$,
where $N_R$ is the few number of unit cells
required to include nonzero overlaps.
On the other hand, the total size of the complex eigenvectors,
which is by far the larger set of data, is exactly
$2\times N_k\times N_b \times N_b\times \mathrm{(size\ of\ a\ real\ number)}$.

\begin{figure}
\includegraphics[width=\columnwidth]{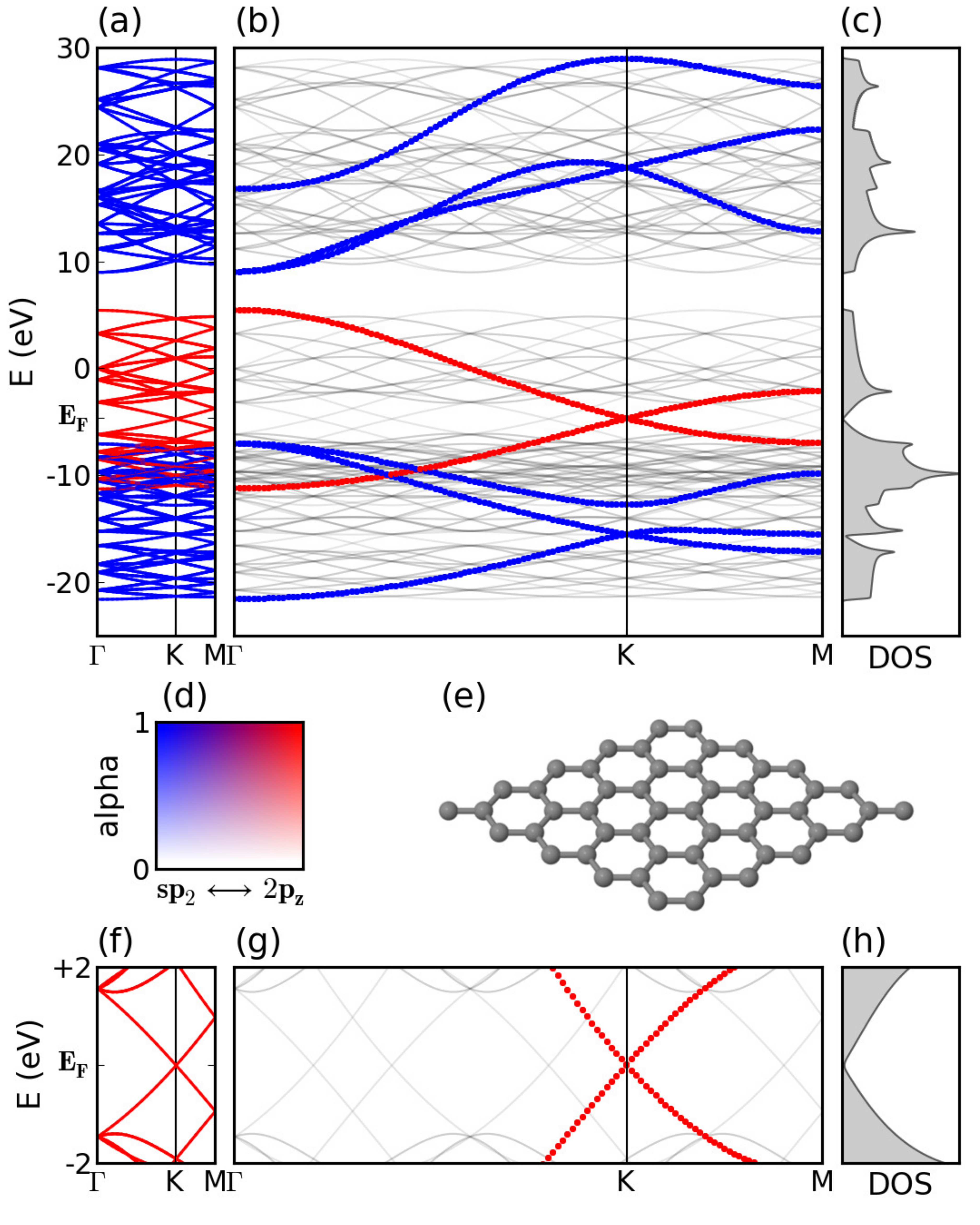}
\caption{\label{fig1} (a) Band structure of graphene
obtained from supercell calculations.
(b) Unfolded bands in the normal Brillouin zone.
The light gray bands are supercell band structure in extended zone.
(c) Total DOS.
(d) Color map used in band structure plots.
(e) The $5\times5$ unit cell used in supercell calculations.
(f--h) Band structure and DOS limited to a window of
$E_F\pm2$~eV centred on the Dirac point.}
\end{figure}

\subsection{Unfolding procedure}
We need an unfolding formula that takes into
account non-orthogonality of atomic orbitals.
Such a formula has been derived recently by
Lee \emph{et al.}\ in Ref.~\onlinecite{lee2013}.
In this section,
we describe their method from the viewpoint of numerical implementation.

It is customary to use upper and lower cases
to refer to supercell and 
normal cell variables, respectively, so $\Kvec$ and $\kvec$ denote
the wavevectors of the supercell Brillouin zone (SBZ)
and normal cell Brillouin zone (NBZ), respectively.
A given $\kvec$ in NBZ folds onto exactly one $\Kvec$ in SBZ but, in contrast,
each $\Kvec$ unfolds back to $N_u$ normal system
$\kvec$ vectors in NBZ.
The unfolding method assigns the same energy at a given $\Kvec$
to the multiple unfolded $\kvec$'s with proper spectral weights.
In practice this simple procedure can be followed.
If we choose a $k$ path in the NBZ,
and regard it as a path in the extended zone of the supercell,
then each $\Kvec$ needs to be unfolded to only a single $\kvec=\Kvec$
in order to achieve the desired band structure representation.
We just calculate the energy eigenvalues along this path for the supercell sytem,
and determine their spectral weights by the unfolding formula.

It is useful to note some properties of non-orthogonal orbitals.
The normalization of an eigenvector, $|\Kvec J\rangle$,
in terms of its expansion coefficients, is given by
\begin{equation} \label{N}
  \sum_{MN} C^{\Kvec J\ast}_N S_{NM}(\Kvec) C^{\Kvec J}_M=1,
\end{equation}
where $S(\Kvec)$ is the overlap matrix of Bloch orbitals,
\begin{equation} \label{S}
  S_{NM}(\Kvec) = \sum_\Rvec
  e^{i\Kvec\bm{\cdot}\Rvec} S_{NM}({\Rvec}).
\end{equation}
Here the overlap integrals are defined by
\begin{equation}
  S_{NM}(\Rvec)\equiv S_{0N,\Rvec M}=\langle 0N | \Rvec M \rangle.
\end{equation}
In Eq.~(\ref{N}), $J$ and $N$ (or $M$) are band and orbital indices,
respectively, and are both in the range of 1 to $N_b$.
Breaking down the contributions in Eq.~(\ref{N}),
we can write the orbital populations for an eigenstate as
\begin{equation} \label{P}
P^{\Kvec J}_N = \Re \left\{
C^{\Kvec J\ast}_N \sum_{M} S_{NM}(\Kvec) C^{\Kvec J}_M \right\}.
\end{equation}

The unfolding procedure must uncover the hidden translational symmetry
inherited by the supercell system from the normal system.
A useful connection between the two systems
is expressed by the Fourier relation,
\cite{allen2013}
\begin{equation} \label{G}
  \frac{1}{N_u}
  \sum_\Gvec e^{i\Gvec\bm{\cdot}\rvec}
  =\delta_{\rvec,\Rvec},
\end{equation}
where $\Rvec$ and $\rvec$ are the corresponding lattice vectors, respectively,
and the sum is over the set of $N_u$ reciprocal lattice vectors
of the supercell system that unfold the SBZ onto the NBZ.
An orbital described in the supercell system by the pair $\Rvec,M$
can be described in the normal system through the mapping,
\cite{lee2013}
\begin{equation} \label{map}
  \Rvec, M \rightarrow \Rvec+\rvec_M, m_M,
\end{equation}
where $\rvec_M$ is one of the $N_u$ normal system
lattice vectors within a supercell.
The range of $m_M$ is the (possibly variable)
number of orbitals in each normal unit cell.

The unfolded spectral weight is given by
\cite{lee2013}
\begin{equation} \label{W}
  W^{\Kvec J}(\Gvec) = \frac{1}{N_u}
  \sum_{MN} C^{\Kvec J\ast}_N U_{NM}(\kvec) C^{\Kvec J}_M,
\end{equation}
where $\kvec=\Kvec+\Gvec$, and
\begin{equation} \label{U}
  U_{NM}(\kvec)=\sum_\rvec
  e^{i\kvec\bm{\cdot}(\rvec-\rvec_M)}
  S_{0N,\rvec m_M},
\end{equation}
which is a particular Fourier sum of overlap integrals
that encapsulates the unfolding information.
We shall call it the \emph{unfolding matrix}.
A convenience of this unfolding formula is that it does not require
the definition of a virtual crystal for its implementation.

Noting the similarity between $U(\kvec)$ and $S(\Kvec)$,
we write partial unfolded orbital weights as
\begin{equation} \label{PW}
  W^{\Kvec J}_N(\Gvec) = \frac{1}{N_u} \Re \left\{
  C^{\Kvec J\ast}_N \sum_{M}U_{NM}(\kvec) C^{\Kvec J}_M \right\},
\end{equation}
so that
$W^{\Kvec J}(\Gvec)=\sum_N W^{\Kvec J}_N(\Gvec)$.
We can verify that the unfolding matrix is hermitean,
so that the spectral weights are properly real.
This can be shown by writing
$S_{0N,\rvec m_M}$ as $S_{\rvec_Nn_N,\rvec m_M}$,
and using the symmetry properties of overlap integrals.
Another checkpoint is the special case of an orthonormal basis set.
We can easily see that orthogonality implies that
$S_{0N,\rvec m_M}=\delta_{\rvec,\rvec_N}\delta_{n_N,m_M}$,
from which we obtain the simpler result,
\cite{ku2010}
\begin{equation}
  U_{NM}(\kvec)=e^{i\kvec\bm{\cdot}(\rvec_N-\rvec_M)} \delta_{n_N,m_M}.
\end{equation}

The main numerical task pertains to the unfolding matrix in Eq.~(\ref{U}).
First, there is a sum over the infinite set of lattice vectors $\rvec$,
but the actual set is sharply restricted by the range of overlap integrals.
Second, it must be noted that the overlap integral
$S_{0N,\rvec m_M}$ is equivalent to one in the standard form of $S_{0N,\Rvec M'}$,
supplied by the electronic structure code.
The supercell and normal cell lattice vectors are related by
$\rvec=\Rvec+\rvec_i$,
where $\rvec_i$ is a normal lattice vector within a supercell,
while the orbital index $M'$ is determined by the pair of $\rvec,m_M$.

Two sum rules can help us verify the results of our calculations.
By using Eq.~\ref{G}, we can prove the first one,
\begin{equation}
  \sum_\Gvec W^{\Kvec J}(\Gvec) = 1.
\end{equation}
The second one, which is more practical for our procedure, is given by
\begin{equation} \label{sr2}
  \sum_J W^{\Kvec J}(\Gvec) = \frac{N_b}{N_u},
\end{equation}
where the right hand side is just the average number of basis orbitals
in the primitive unit cell of the normal system.
This sum rule is a consequence of the properties of the spectral function.

\begin{figure}
\includegraphics[width=\columnwidth]{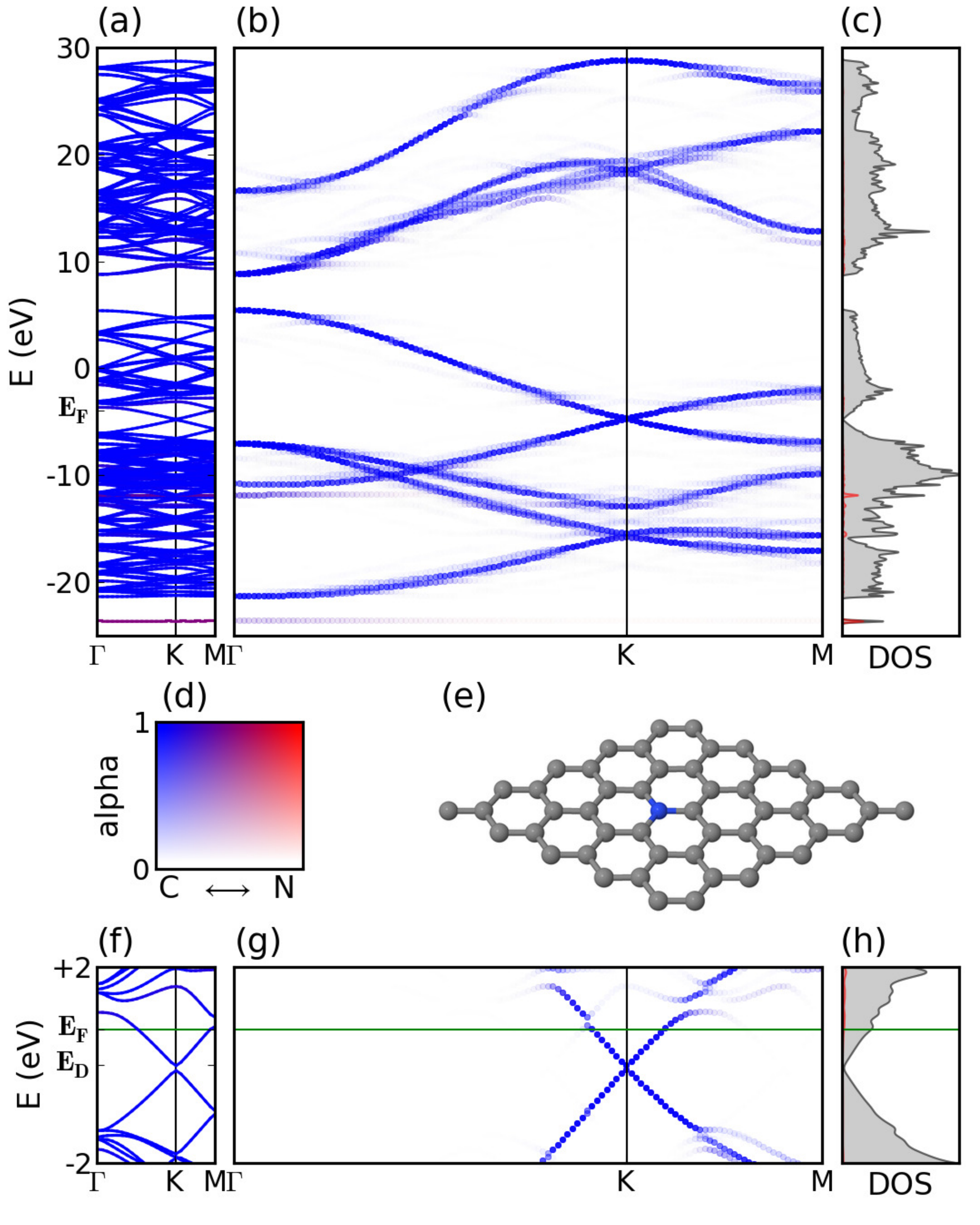}
\caption{\label{fig2} (a) Band structure of nitrogen-doped graphene
obtained from supercell calculations.
(b) Unfolded bands in the normal Brillouin zone.
(c) Total DOS and PDOS of nitrogen orbitals.
(d) Color map used in band structure plots.
(e) The $5\times5$ unit cell with a nitrogen at the centre.
(f--h) Band structure and DOS plotted in a window centred on the Dirac point.
Fermi level is seen to have shifted because of doping by the extra electron of nitrogen.
There is a small band gap opening at the Dirac point.}
\end{figure}

\section{Results and discussion}
\subsection{Perfect graphene} \label{perfect}
Band structure of perfect graphene provides a good testing ground
for our numerical procedure,
since we know exactly what to expect.
Furthermore, it allows us to measure the computational load
of the unfolding procedure,
which is about the same in all cases that we consider.
Using a Fortran serial code,
on a computer with
\textsf{Intel\textregistered}\ \textsf{Core}$^\mathsf{TM}$
\textsf{i7-3610QM} CPU,
the runtime for the unfolding of this example was about one minute.

We first explain our approach to using
additional data consisting of
unfolded spectral weights and orbital populations
in the visualization of band structures.
(A somewhat different approach is described in the \textit{Supplemental Material} of
Ref.~\onlinecite{berlijn2011}.)
In the (\textsf{matplotlib}) 2D plotting package, used here,
we can specify colors and transparency in terms of parameters $r,g,b$ and $\alpha$,
respectively, all in the 0--1 range.
\cite{hunter2007}
Hence in the $rgb$ color space (0,0,0) and (1,1,1) represent black and white,
respectively, which are visible or invisible on a white background.
The $\alpha$ parameter controls transparency,
so that $\alpha=0$ is completely transparent and thus invisible,
and $\alpha=1$ is completely opaque.
We adopt a simple scheme and
always decompose the orbital contributions into only two sets,
which are represented by blue (0,0,1) and red (1,0,0), respectively.
We then use the normalized weighted average of
the colors to represent mixing of the two sets,
and set $\alpha$ equal to the total spectral weight,
which results in the color map shown in Fig.~\ref{fig1}(d).
It is possible to partition the orbitals into more than
two sets but the color maps become more complicated.
We must mention that this continuous range of colors is unnecessary for
the present case of perfect graphene,
where the weights turn out to be zero or unity,
but it will be essential for defective graphene.

Figure~\ref{fig1}(a) shows the band structure,
using 25 $k$ points along $\Gamma$KM path in the SBZ.
We have assigned our two pure colors to the two sets
consisting of $2s$, $2p_x$, $2p_y$ orbitals
and the $2p_z$ orbital, respectively.
This has made the $\sigma$ and $\pi$ bands
to appear in these colors,
since these bands are mutually orthogonal in graphene.
However, heavy folding makes it hard to recognize the shape of the bands.

Figure~\ref{fig1}(b) shows the unfolded bands.
We make the calculations using the same $5\times5$ unit cell
along the $\Gamma$KM path in the NBZ of graphene.
As this path is five times larger than the one in SBZ,
we use 121 $k$ points.
The expected bands of graphene have emerged, in blue and red,
out of the complicated supercell bands, shown in light grey.
Figures~\ref{fig1}(f--h) show the same data as in Figs.~\ref{fig1}(a--c),
but in the important range of low energies.

\begin{figure}
\includegraphics[width=\columnwidth]{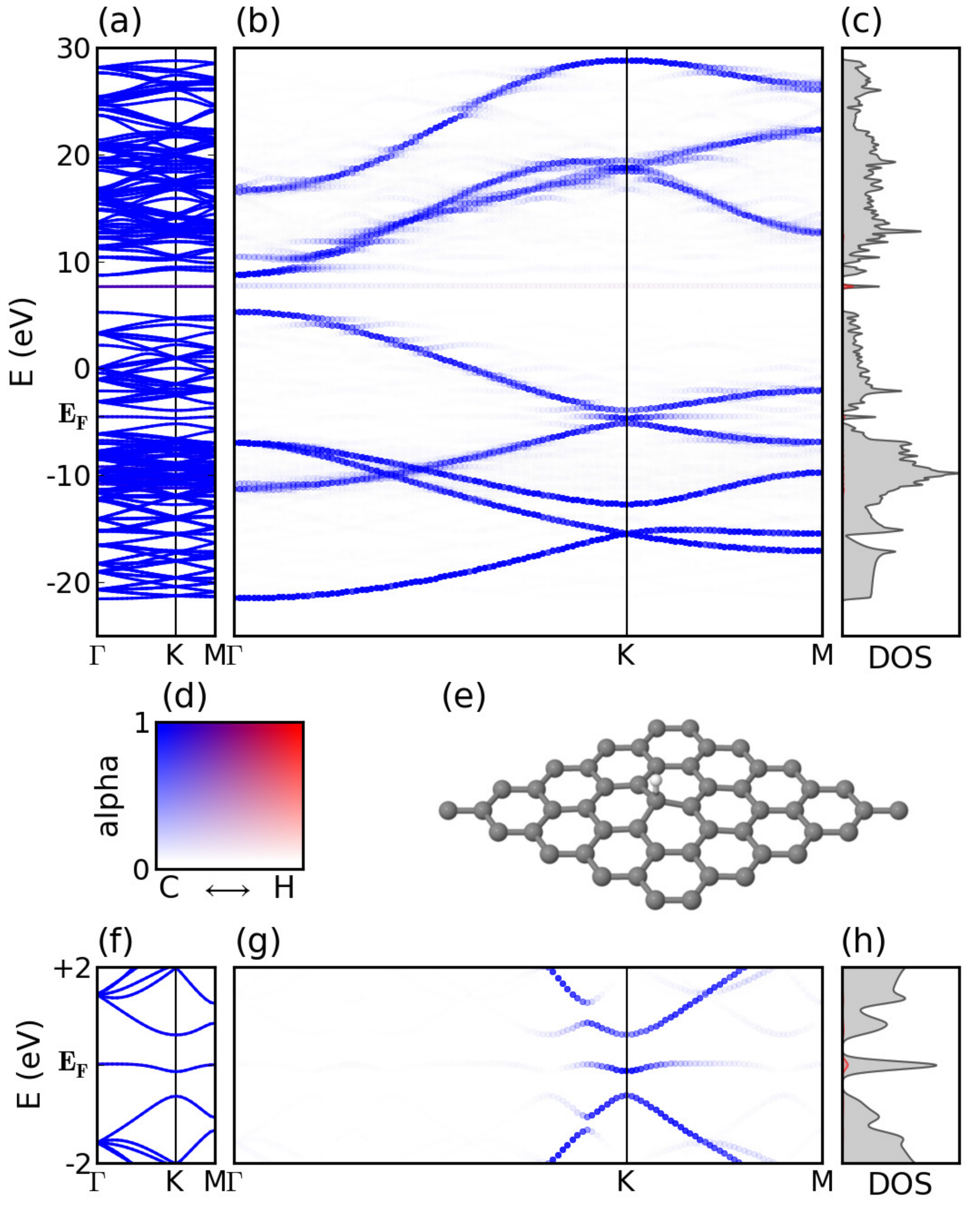}
\caption{\label{fig3} (a) Band structure of graphene
with a hydrogen adsorbate obtained from supercell calculations.
(b) Unfolded bands in the normal Brillouin zone.
(c) Total DOS and PDOS of hydrogen orbital.
(d) Color map used in band structure plots.
(e) The $5\times5$ unit cell with a hydrogen at its centre.
A buckling of the lattice is visible near H.
(f--h) Band structures and DOS plotted
in a 4 eV window centred on the Dirac point.}
\end{figure}

\subsection{Nitrogen impurity}
We repeated similar electronic structure calculations for a $5\times5$ graphene
supercell with a single nitrogen substitutional impurity.
The results are shown in Fig.~\ref{fig2},
where we have used the same visualization approach discussed in Sec.~\ref{perfect},
except that we have partitioned the orbital contributions
according to the atoms, C and N.
The utility of a continuous color map is demonstrated
by the existence of bands with contributions from both C and N orbitals.
We note the appearance of the bound states below
each of $\sigma$ and $\pi$ bands,
because of the presence of nitrogen impurity.
\cite{lambin2012}
In addition, unfolding shows
that the weights of these states have nonuniform distributions
in the NBZ.

In Figs.~\ref{fig2}(f--h) we show the data for the low energy spectrum.
Because of doping by nitrogen, the Fermi level has shifted above the neutrality point.
The presence of impurity has also caused a small gap to open at the Dirac energy.

\subsection{Hydrogen adsorbate} \label{hydrogen}
Results of calculations for a single hydrogen adsorbate
on a $5\times5$ graphene supercell are shown in Fig.~\ref{fig3}.
A buckling of the graphene lattice
near hydrogen, seen in Fig.~\ref{fig3}(e),
indicates the formation of an $sp^3$ defect.
Such a defect effectively creates a $\pi$ orbital vacancy,
which manifests itself as a midgap state at the Fermi level.
\cite{wehling2009,mirzadeh2012}
We can see the midgap state in all the band structure plots
as well as the DOS plots.
Moreover, most of the weight of the midgap band comes from carbon
$2p_z$ orbitals.
There is a bound state visible in the gap above the $\pi$ bands
with contributions from both C and H atoms, Figs.~\ref{fig3}(a--c).
\cite{farjam2011}
We note, in Fig.~\ref{fig3}(g), that the weight of the midgap
band is accumulated near the $K$ point,
a fact that has also been observed in ARPES spectra of hydrogenated graphene.
\cite{haberer2011}

\subsection{Vacancy defect}
\begin{figure}
\includegraphics[width=\columnwidth]{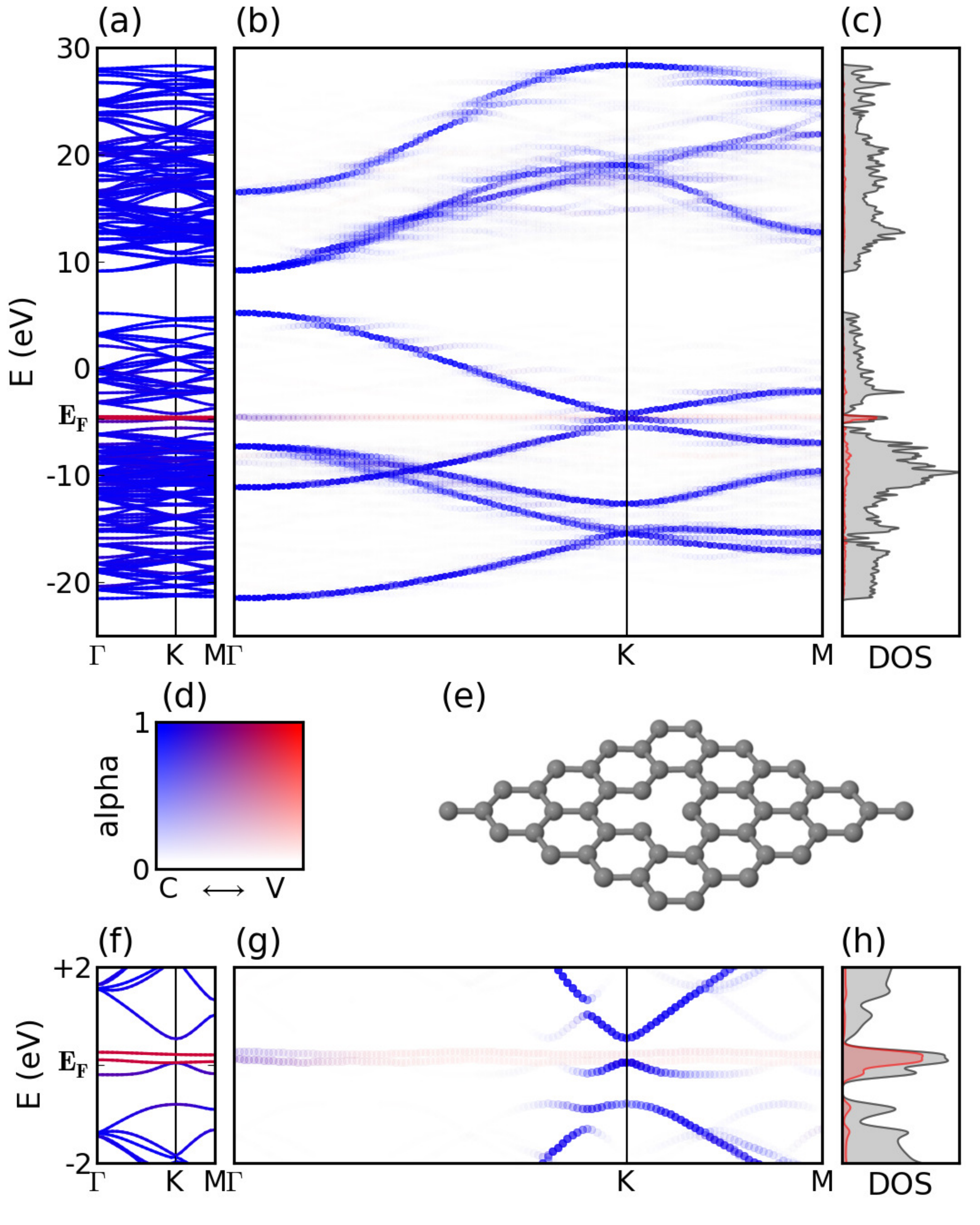}
\caption{\label{fig4} (a) Band structure of graphene
with a vacancy obtained from supercell calculations.
(b) Unfolded bands in normal Brillouin zone.
(c) Total DOS and the PDOS of three carbon atoms bordering the vacancy site.
(d) Color map used in band structure plots.
(e) The $5\times5$ unit cell with a vacancy at its centre.
(f--h) Band structures and DOS in a 4 eV window centred on the Dirac point.}
\end{figure}

Electronic structure calculation for graphene with a missing atom
is shown in Fig.~\ref{fig4}.
Although not visible in Fig.~\ref{fig4}(e),
the relaxation of atoms results in a Jahn-Teller distortion.
\cite{yazyev2007}
There are four missing orbitals at the vacancy site,
which can give rise to midgap bands,
and other spectral changes.
\cite{nanda2012}
Three midgap bands are clearly displayed in Figs.~\ref{fig4}(f--h),
where the use of the color map
has also revealed the difference in their orbital decomposition.
The two pure colors were assigned to two sets of carbon atoms,
those nearest the vacancy site and those on the remaining sites.

\begin{figure}
\includegraphics[width=\columnwidth]{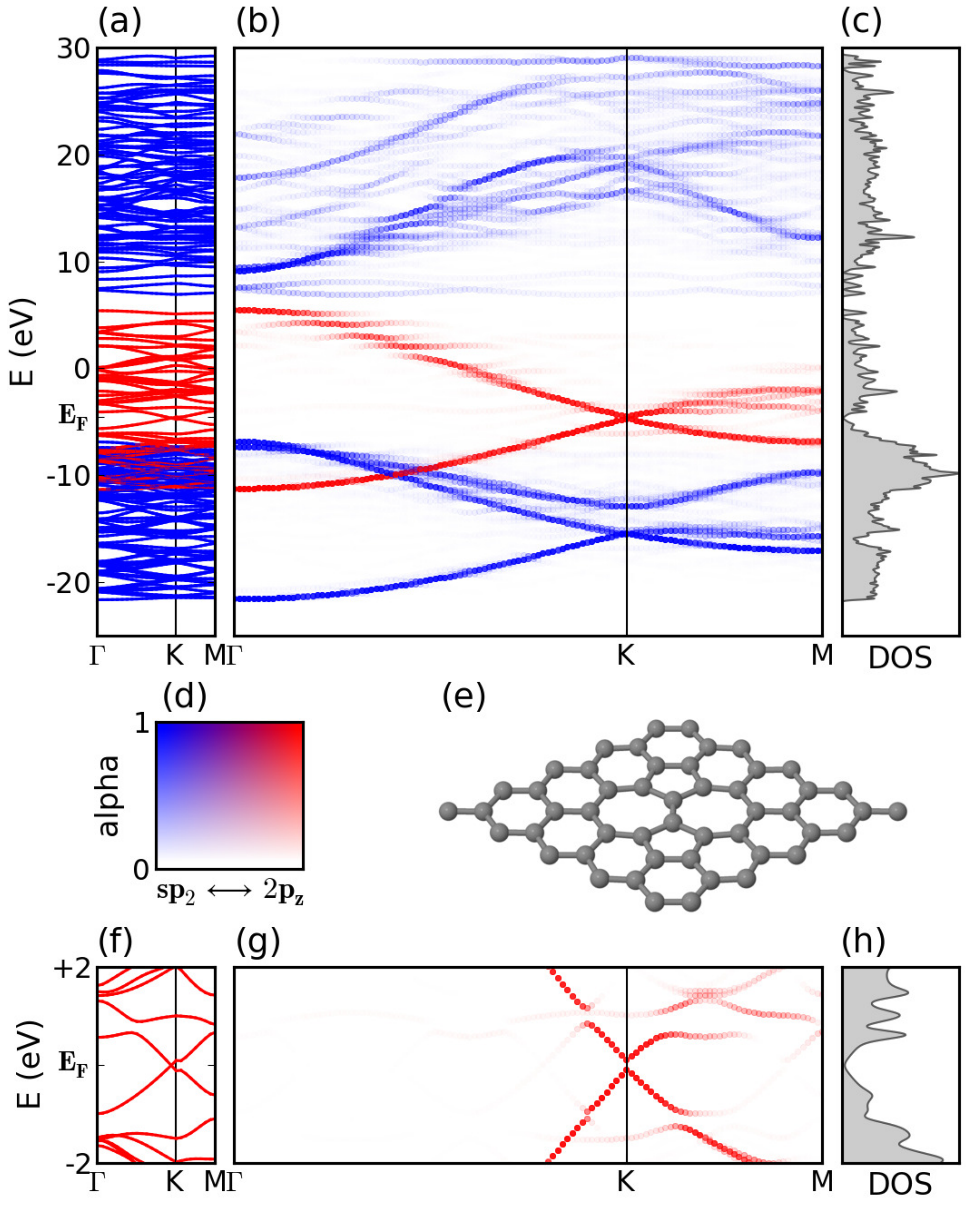}
\caption{\label{fig5} (a) Band structure of graphene
with a Stone-Wales defect obtained from supercell calculations.
(b) Unfolded bands in normal Brillouin zone.
(c) Total DOS.
(d) Color map used in band structure plots.
(e) The $5\times5$ unit cell with the Stone-Wales defect.
(f--h) Band structures and DOS in a 4 eV window centred on the Dirac point.}
\end{figure}

\subsection{Stone-Wales defect}
Lastly, we performed calculations for the SW defect.
A similar example was recently treated with a different band unfolding
methodology in Ref.~\onlinecite{medeiros2014},
which could provide an interesting comparison.
Although the Dirac cone is generally preserved as pointed out in
Ref.~\onlinecite{medeiros2014},
a small gap has appeared at the $K$ point of the unfolded bands,
shown in Fig.~\ref{fig5}(g).
A close examination of the folded bands, Fig.~\ref{fig5}(f),
shows that there is actually no gap in the energy spectrum,
but the band crossing has slightly shifted from its SBZ $K$ point
toward the $\Gamma$ point.
Also, the SW defect has caused a lot of smearing and breaking up of the bands,
and appearance of new bands,
especially for the $\sigma$ bands.
It can be expected that the magnitude of the defect-induced effects will
diminish smoothly as the size of the supercell is gradually increased.

In all cases, we checked that the sum rule, (\ref{sr2}), holds.
For pure graphene, SW defect or nitrogen impurity,
the sum adds up to $N_b/N_u=200/25=8$,
while for hydrogen and vacancy defects it adds up to $201/25=8.04$,
and $196/25=7.84$, respectively.
The variety of the examples used show the versatility of our procedure.

\section{Conclusions}
We have implemented the unfolding method of Lee \emph{et al.},
\cite{lee2013}
have proposed how to visualize additional data from spectral weights
in energy dispersion plots,
and have demonstrated these for selected defective 2D graphenes.
Although the DFTB+ code was used in this work,
the unfolding procedure described can more generally be employed for any
electronic structure code that uses a basis of non-orthogonal
atomic orbitals.
It can be predicted that a great number of graphene-based systems,
involving defects, impurities, adsorbates and substrates
can take advantage of the methods discussed here.
The simplified effective band structures could be more easily
comprehended,
and supercell calculations may be compared directly with normal cell
calculations.
The major attraction, however, is the correlation of
the unfolded band structure
with the measurements of angle-resolved photoemission spectroscopy
(ARPES) experiments.

\begin{acknowledgments}
I have greatly benefited from discussions with Alex~Gr\"uneis.
\end{acknowledgments}

\bibliography{main}

\end{document}